\documentclass[12pt,preprint]{aastex}

\begin{document}

\title{Three New Long Period X-ray Pulsars Discovered in the Small  
Magellanic Cloud} 

\author{D.J. Macomb}
\affil{Physics Department, Boise State University, Boise, ID 83725-1570}
\email{dmacomb@boisestate.edu}

\author{D. W. Fox, R.C.Lamb, T.A. Prince}
\affil{Space Radiation Laboratory, California Institute of Technology,
Pasadena, CA 91125}
\email{derekfox@astro.caltech.edu,lamb@srl.caltech.edu,prince@srl.caltech.edu}

\begin{abstract}
The Small Magellanic Cloud is increasingly an invaluable laboratory
for studying accreting and isolated X-ray pulsars.  We add to the
class of compact SMC objects by reporting the discovery of three new
long period X-ray pulsars detected with the {\it Chandra X-ray
Observatory}.  The pulsars, with periods of 152, 304 and 565 seconds,
all show hard X-ray spectra over the range from 0.6 - 7.5
keV.  The source positions of the three pulsars are consistent with
known H-alpha emission sources, indicating they are likely to be Be
type X-ray binary star systems.
\end{abstract}


\section{Introduction}

Our program of routinely evaluating the pulsar content of archival
observations of the Small Magellanic Cloud has previously revealed
five new X-ray pulsars (Macomb et al. 1999, Finger et al. 2001, Lamb
et al. 2002a, Lamb et al. 2002b).  To date, there are at least 31
known X-ray pulsars in the Small Magellanic Cloud, 25 of which were
discovered in the last six years (Yokogawa 2002, Lamb et al. 2002a,
Laycock et al. 2002, Corbet et al. 2002, Lamb et al. 2002b).  This
explosion in the number of known pulsars is due mainly to four X-ray
missions, ASCA, ROSAT, RXTE, and Chandra, each of which has devoted time
to long, sensitive observations of the SMC.

The advantages to studying the compact object content of the
Magellanic Clouds are well known: a relatively small angular size, at
a reasonably close and known distance, located at a high galactic
latitude with little obscuration by the interstellar medium of the
Galaxy.  In addition, there is evidence for recent star formation in
the SMC over the last few 10's of millions of years which creates an
environment in which X-ray pulsars are expected to be plentiful
(Haberl \& Sasaki 2000, Maragoudaki et al. 2001).

Our analysis of a recent Chandra X-ray Observatory observation of the
SMC reveals evidence for four X-ray pulsars.  One, CXOU
J011043.1-721134, is possibly an anomalous X-ray pulsar (Lamb et
al. 2002b).  This paper describes the discovery of X-ray pulsations
from the other three sources: CXOU J005750.3-720756 (152.10 s
period), CXOU J010102.7-720658 (304.49 s period) and CXOU
J005736.2-721934 (564.83 s period).  After a description of the
discoveries, we address the characteristics of each source,
followed by a discussion of the general significance of these
detections.

\clearpage
\begin{deluxetable}{lccll}
\tabletypesize{\scriptsize}
\tablecolumns{5} 

\tablecaption{Names and positional information for the three X-ray
pulsars. Errors are in parenthesis for each column.}

\tablewidth{0pt}

\tablehead{\colhead{Source} & \colhead{Right Ascension (J2000)\tablenotemark{a}} & \colhead{Declination (J2000)\tablenotemark{a}} & \colhead{Proposed counterpart\tablenotemark{b}} & \colhead{Counterpart Offset (arcsec)} }

\startdata

CXOU J005750.3-720756 & 14.45975(2)& -72.13239(1)& [MA93] 1038 &0.67\\
CXOU J010102.7-720658 & 15.26147(19)& -72.11614(6)& [MA93] 1240& 1.14 \\
CXOU J005736.2-721934 & 14.40123(10)& -72.32637(3)& [MA93] 1020 & 1.64\\

\enddata

\tablenotetext{a}{statistical error quoted; total systematic error on the position could be as much as 1 arcsecond}
\tablenotetext{b}{Source number from the catalog of Meyssonnier \& Azzopardi 1993}
\end{deluxetable}
\clearpage

\section{Discovery of Pulsations}

The discovery data were obtained from CHANDRA obsid 1881, a 100ks
ACIS-I observation which began 2001 May 15.  The ACIS-I read-out time
for this observation was set at a coarse 3.241 seconds, giving a
Nyquist frequency of 0.154 Hz.  Processing of the ACIS field reveals
140 point sources detected at the 3 sigma level using a standard
sliding box detection algorithm.  We extracted photons for each point
source in the field using an ellipse of varying size, defined by the
source detection software, which encompassed $90\%$ of the source
photons.  

The photon arrival times were barycentered and a fast Fourier
transform performed for each individual source.  With the average fft
power normalized to one, we set a power threshold of 20 to flag
potential signals.  For a 100 ksec observation binned at 3.241
seconds, there are about 31000 independent trials.  This gives a
probability of around $1\%$ that any peak power beyond 20 arises by
chance for all of the 140 sources analyzed.

Concentrating on frequencies above 0.001 Hz (a 1000 second period), we
find 21 of the 140 sources with powers beyond 20.  Of these, all but
four sources had powers at two set frequencies that appeared in
multiple sources.  The first frequency was near 0.001 Hz and the
second at 0.001415 Hz, periods of 1000 and 706 seconds, both of
which are attributable to known dithering effects (as described at
http://cxc.harvard.edu/ciao/why/dither.html).  Of the four
remaining sources, one is CXOU J011043.1-721134 which is a 5.44 s
pulsar (Lamb et al. 2002b), leaving us with three new candidate pulsars.
Table 1 lists the Chandra centroid for each source along with the
proposed counterpart. 

The Fourier power distributions for the three remaining sources are
shown in Figure 1.  These pulsar timing distributions are based upon a
slightly different photon set than for the general search.  These
refined photon selections encompassing the energy range from 0.6 - 7.5
keV are close to maximizing the Fourier power for all
three sources.  In addition, we chose position regions corresponding
to an 80\% encircled energy to further optimize pulsar timing
parameters.  These selections resulted in 5429 photons
for the 152 s pulsar, 325 photons for the 304 second pulsar, and 3141
photons for the 565 second pulsar.

The first source, centered on a frequency of 0.0065747 Hz (152.10
seconds) has a peak power of 90 while the second is at a frequency of
0.0032842 Hz (304.49 seconds) with peak power 30.4.  While this
frequency is close to being twice the frequency of the 152 s pulsar,
they are not precisely doubled.  In addition, the fact these sources
are in different ACIS chips and that neither of these frequencies show
up in any other field source no matter the source strength indicates
that they are in fact distinct signals.  The third source is detected
at a power of 57 at a frequency of 0.00177045 Hz (564.83).

In the case of the 152 and 565 second pulsars, a third harmonic of the
fundamental frequency is clearly seen, and there is also a strong
second harmonic for the 152 second pulsar.  No other sources in the
field of view have significant Fourier power at any of these
frequencies.  The implication is that these three sources are
long-period X-ray pulsars.  As their individual characteristics make
clear, they are both likely to be examples of high-mass X-ray binary
systems with Be star companions.

There is little evidence of variability from any of the three new
sources over the 100 ksec {\it Chandra} observation.  While a $\chi^{2}$
test of variability for the pulsars binned over single pulse periods
for the 152 and 565 second pulsars and 5 pulse periods for the 304
second pulsar shows little gross variability, there is modest
variability for the 565 second pulsar (3\% chance of being consistent
with a flat rate).  There is no evidence for any of the pulsars
dramatically changing their flux during the observation, i.e. an onset
or end of an emission episode. Consequently, we have taken all
spectral and timing information for these three pulsars from the full
observation.
\clearpage
\begin{deluxetable}{llllllll}

\tablecolumns{8} 

\tablecaption{Pulsar characteristics and power-law spectral fit
information for the three X-ray pulsars.  The luminosity values are
for the energy interval 0.6 - 7.5 keV.}

\tablewidth{0pt}

\tablehead{\colhead{Source} & \colhead{period(s)} & \colhead{pulse fraction} & \colhead{n$_{H}$\tablenotemark{a}} & \colhead{$\alpha$} & \colhead{norm\tablenotemark{b}} & \colhead{$\chi^{2}$ (dof)} &\colhead{L\tablenotemark{c}} }

\startdata

005750.3-720756 &152.098(16) & $64\pm3\%$ & $0.57\pm0.05$ &$1.01\pm0.06$ &$7.2\pm1.1$ & 0.98 (227) & $26$ \\
010102.7-720658& 304.49(13) & $90\pm8\%$ & 0.66\tablenotemark{d}& $1.22\pm0.26$ & $0.4\pm0.2$ &0.47 (13) & $1.1$\\
005736.2-721934 & 564.81(41) & $48\pm5\%$ &$0.76\pm0.14$ &$1.70\pm0.23$ &$4.1\pm2.1$ &0.47 (134) & $5.6$ \\

\enddata

\tablenotetext{a}{units of 10$^{22}$ cm$^{2}$}
\tablenotetext{b}{normalization at 1 keV in units of 10$^{-5}$ photons cm$^{-2}$ s$^{-1}$}
\tablenotetext{c}{Unabsorbed luminosity assuming a distance of 57 kpc in units of 10$^{34}$ ergs/s}
\tablenotetext{d}{n$_{H}$ value frozen at 0.66, the mean of n$_{H}$ for the other two pulsars, otherwise unconstrained}
\end{deluxetable}
\clearpage
\section{Individual Source Characteristics}

\subsection{The 152 second pulsar: CXOU J005750.3-720756}

A search of the ROSAT observation catalog shows many ROSAT HRI or PSPC
observations overlapping the position of CXOU J005750.3-720756.  In
addition, the ROSAT results archive lists a source, RX J0057.8-7207,
a source with an 11 arcsecond error radius that is within 6 arcsecond 
of the Chandra position centroid.  No previous
reports of X-ray pulsations from this source have been reported,
however it was suggested as a Be binary pulsar candidate by Haberl \&
Sasaki\ (2000) based upon positional coincidence with an H-alpha
region cataloged by Meyssonnier \& Azzopardi (1993).  Given the small
positional difference between the CHANDRA position (absolute position
good to approximately 0.6'') and the centroid of the ROSAT/PSPC source
(localization 5.1'', Haberl \& Sasaki 2000), it is likely that these
two sources are the same. Unfortunately, a check of the three
strongest ROSAT detections of this source does not reveal evidence for
pulsed emission. 

This is not surprising, however, when one considers the energy
spectrum of this source.  The spectrum of CXOU J005750.3-720756 has a
hard photon index of $1.0\pm0.1$.  The complete power law
fit parameters are listed in Table 2, along with the estimated pulse
period and period error.  The error on the pulse period is calculated 
from the Rayleigh power, based upon the frequency points corresponding
to a drop in power equal to the square root of the peak power.
The source flux is the highest
of these three sources.  Given this very hard spectrum, it is
likely that a soft X-ray telescope with reduced effective area
such as the ROSAT/PSPC and
ROSAT/HRI are less effective at detecting the periodicity.

Figure 2 shows the folded Chandra lightcurve for the 152.1 second
pulsar.  The pulse profile is complicated, showing a sharp dip with a
duration of 10\% of the pulse period.  The pulsed fraction is
63$\pm3$\%.  A search of the full dataset over frequency derivative
(binning at 1.0x10$^{-13}$ Hz/s) and frequency provides a small but
statistically insignificant improvement in detected power.  This is
the case for all three pulsars, so all quoted pulse periods and
frequencies are uncorrected for frequency derivative.

\subsection{The 304 second pulsar:  CXOU J010102.7-720658}

The weakest of the three sources by an order of magnitude, CXOU
J010102.7-720658 has a less well-defined pulse profile as shown by
Figure 2.  The source has a pulsed fraction of 90$\pm$8\%, a
fortuitously large value since if the pulsed fraction were more like
the other two, we would not have detected the periodicity.  This
source, with only 470 total photons considered for
spectroscopy, has a spectral shape which is more difficult to
quantify.  Table 2 lists the power-law fit parameters.  The spectrum
is slightly softer than the 152 second pulsar, but the power-law
spectral shape is not the best fit.  A blackbody fit gives a
temperature of 1.0 keV, consistent with it being a hard spectrum.

As with the 152 second pulsar, this previously known X-ray emitter was
postulated to be an X-ray binary pulsar by Haberl \& Sasaki and by
Kahabka \& Pietsch (1996) based upon correlations of the ROSAT X-ray
source, RX J0101.0-7206, with an H-alpha emission source.  The
putative ROSAT counterpart is only 0.96 arcseconds from the CHANDRA
detection, well within the combined source location errors for the two
detectors indicating they are the same source.  The most remarkable
thing about this source is the extremely weak flux.  Assuming a
distance of 57 kpc for the SMC, the X-ray luminosity is only
1.1x10$^{34}$ ergs/s. This is a very small discovery flux, making it
the weakest accreting SMC source for which X-ray pulsations have been
discovered.

\subsection{The 565 second pulsar: CXOU J005736.2-721934}

The pulse profile of CXOU J005736.2-721934 is shown in Figure 2.  The
pulse profile is quite complicated, showing three distinct peaks, a
characteristic indicated by the strong presence of the first three
harmonics in the Fourier transform spectrum.  This source is also formally the
most variable of the three with a $\chi^{2}$ test for variability
based upon binning the flux over a single pulse period shows that the
flux is consistent with being constant at the 3\% level unlike the
other two sources, which are fully consistent with constant flux.  
This is not, however, conclusive evidence for pulse to pulse variability.

Unlike the previous two sources, there has been no previous suggestion
that this is an accreting X-ray pulsar.  This source is only 18
arcseconds away, however, from source 19 of the ASCA SMC source
catalog, well within the half-power radius of the ASCA XRT (Yokogawa
et al. 2000). Its pulsed fraction is 48$\pm$5\%, and it is has the
softest spectrum of the three sources.  There are no reported ROSAT
detections of this source.  A search of stellar databases does show
that there is an H-alpha emission line source located within 0.8
arcsecond of the CHANDRA position (Meyssonnier \& Azzopardi - Star 1020
with a J2000 location of 00 57 36.0 -72 19 34.  This lends support to
the interpretation of this source as yet another SMC Be-binary.

\section{DISCUSSION}

There are two independent arguments that these pulsars have high mass
companions.  The first is their apparent association with H-alpha
sources which are themselves markers of regions containing massive
stars (Kennicutt 1983).

We have assessed the possibility that the apparent association of
these 3 pulsars with known H-alpha sources is due to an accidental
overlap in position.  A cross-check of the {\it Chandra} positions with
the H-alpha catalogs of Meyssonnier \& Azzopardi (1996) and Murphy \&
Bessell (2000) shows that 
of the 140 {\it Chandra} sources in our field, only 4 were
within 5 arcseconds of an H-alpha source.  The three closest
coincidences were for these 3 pulsars - CXOU J005736.2-721934 (565 s)
at a distance of 1.64'', CXOU J010102.7-720658 (304 s) at a distance of
1.14'', and CXOU J005750.3-720756 (152 s) at a distance of 0.67''.  The
binomial probability of these three particular sources having the
smallest 3 distances to an H-alpha source is $10^{-6}$.  Meyssonnier \&
Azzopardi do not quote precise error radii for their sources, but
indicate that overall ``seeing'' for the observations was on the order
of 1 arcsec.  In addition, the CHANDRA positions are expected to be good
to within about 0.6 arcsec.  
Thus, the distances for all three pulsars from the putative Halpha
source are within a sum
of the estimated {\it Chandra} and Meyssonnier \& Azzopardi error radii
($\sim$1.6'').  

The second argument which supports the identification of these pulsars
as high mass systems is given by the empirical correlation noted by
Corbet (1986) which, for X-ray pulsars with spin periods in excess of
100 s, implies a companion which is either a Supergiant or a Be
star.

Thus, like virtually all other SMC pulsars (31 total), these three are
associated with high mass binaries.  In contrast, from the Milky Way
with approximately 100 times the mass of the SMC only about 40 such
objects have been detected.  This overabundance has been discussed in
some detail by Schmidtke et al. 1999 and Yokogawa et al. 2000.  This
overabundance points to a rather recent star formation activity $\sim$
10$^7$ yr ago.

Further evidence for this episode of star burst activity comes from
examination of the spatial distribution of the SMC X-ray binaries.
Maragoudaki et al. (2001) have recently analyzed the star content of
the SMC using UV photometry over a large (6$^\circ$ by 6$^\circ$)
field.  The distribution of the SMC X-ray binaries follows very
closely (with 3 exceptions) the their distribution of the youngest
stars, those having ages less than 8$\times$10$^6$ yr (figure 9,
Maragoudaki et al.)  This distribution differs considerably from the
general distribution of 517 SMC X-ray sources catalogued by Haberl et
al. (2000).  For example, 29\% of the X-ray sources (149 of 517) fall
in the south-eastern region of the SMC (RA greater than 01$^h$ and
Dec. less than -72.5$^\circ$).  On the other hand only 10\% (3 of 31)
of the X-ray pulsars are located in this region.  This is another reflection
of the fact that the population of the SMC pulsars are prominently (if
not, exclusively) high mass and therefore generally younger then other
types of X-ray sources.  Further detailed comparisons between the star
maps of Maragoudaki et al. and the distribution of the SMC X-ray
binaries may provide useful constraints on evolutionary scenarios of
high mass X-ray binaries.

Despite the relatively poor timing resolution (3.241 s) of this
particular observation, the single 100 ksec exposure reveals four new
X-ray pulsars, increasing the number of known SMC pulsars by over
10\%.  A longer exposure with the {\it Chandra} instruments at much
higher time resolution would probably provide even more discoveries
and allow important assessments of the underlying distributions of SMC
pulsars.

\acknowledgements

This work made use of software and data provided by the High-Energy
Astrophysics Archival Research Center (HEASARC) located at Goddard
Space Flight Center.  Additional use was made of the SIMBAD catalogue
at CDS, Strasbourg France, and NASA's Astrophysics Data System (ADS) .

\clearpage

\begin{figure}
\figurenum{1} \epsscale{0.65} 
\plotone{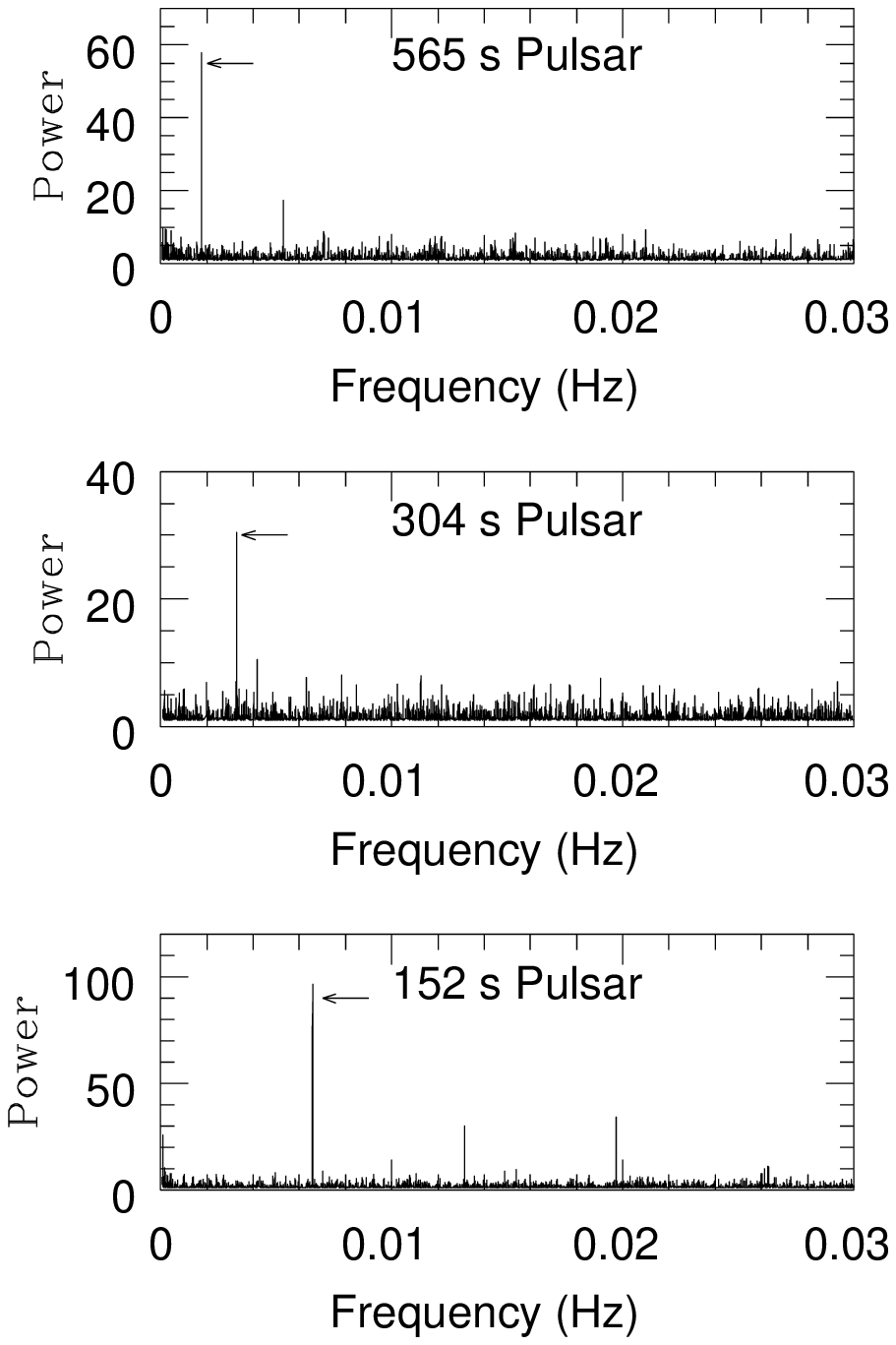}

\caption{FFTs of the CHANDRA sources CXOU J005750.0-720755 (152.10 s
period), CXOU J010102.4-720657 (304.49 s), and CXOU J005735.8-721933
(564.83 s).  The arrow marks the fundamental frequency for each
source.}

\end{figure}

\begin{figure}
\figurenum{2} \epsscale{0.8} 
\plotone{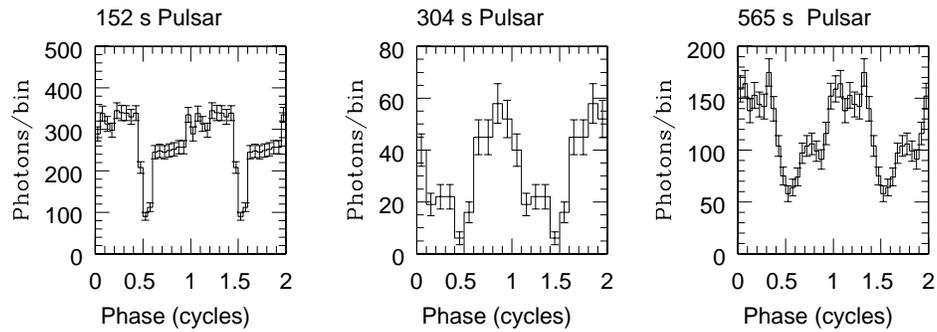}

\caption{Pulse profiles for the three new X-ray pulsars with an energy
restriction of 0.6 to 7.5 keV. The weak, 304 second pulsar only
reveals a single pulse, while the other two have more harmonic
content.  The dashed line in each case shows the off-source background
level.}

\end{figure}

\end{document}